\documentstyle[aps,psfig]{revtex}
\begin{document}
\draft
\title{Microcanonical Approach for the OLA model}
\author{L. Velazquez\thanks{%
luisberis@geo.upr.edu.cu}}
\address{Departamento de F\'{i}sica, Universidad de Pinar del R\'{i}o\\
Mart\'{i} 270, esq. 27 de Noviembre, Pinar del R\'{i}o, Cuba. }
\author{F. Guzm\'{a}n\thanks{%
guzman@info.isctn.edu.cu}}
\address{Departamento de F\'{i}sica Nuclear\\
Instituto Superior de Ciencias y Tecnolog\'{i}as Nucleares\\
Quinta de los Molinos. Ave Carlos III y Luaces, Plaza\\
Ciudad de La Habana, Cuba.}
\date{\today}
\maketitle

\begin{abstract}
In the present paper it is analyzed a very simple example of pseudoextensive
system, the tridimensional system of Linear Coupled Oscillators{\bf \ (OLA
Model). }The same one constitutes a classical tridimensional system of
identical interacting particles by means of harmonic oscillators. This
academic problem possesses a complete analytical solution allowing this way
that it can find application in modeling some properties of the
self-gravitating systems. It is shown that although this is a nonextensive
system in the usual sense, it can be dealt in the thermodynamic limit with
the usual Boltzmann-Gibbs' Statistics with an appropriate selection of the
representation of the space of the integrals of motion.

Keywords: microcanonical ensemble, nonextensive systems
\end{abstract}

\pacs{PACS numbers: 05.20.Gg; 05.20.-y}

\section{Introduction}

In our previous works it was established a general methodology to deal with
some nonextensive systems. In the ref.\cite{vel1} it was addressed the
problem of generalizing the extensive postulates of the traditional
Thermodynamics in order to extend the application of this theory to the
study of some Hamiltonian nonextensive systems. According to our
proposition, This can be performed taking into account the self-similarity
scaling properties of the systems with the realization of the thermodynamic
limit, the limit of many particles, and analyzing the necessary condition
for the equivalence of the microcanonical ensemble with the generalized
canonical one. In the ref.\cite{vel2} it was analyzed the most familiar
self-similarity scaling laws of the systems,{\em \ the exponential scaling}.
Systems with this kind of scaling behavior in the thermodynamic limit can be
dealt with the usual Boltzmann-Gibbs' Statistics if an appropriate selection
of the representation of the space of the integrals of movement, $\Im _{N}$,
is taken. That is the reason why we refer those systems as pseudoextensive.
The extensive systems are just a particular case of the pseudoextensive
systems. It is easy to show that it is sufficient the presence of an
additive kinetic part in the system Hamiltonian for the consideration of
that system as pseudoextensive.

The present paper will be devoted to the microcanonical analysis of a very
simple nonextensive system that allows us a complete analytical study: {\bf %
the tridimensional system of Linear Coupled Oscillators (OLA model)}. This
academic model has been used in the modelation of many systems, and although
it is very well-known from the begining of Mechanics, nowadays it is still
considered in the description of some real system, e.i.: quantum dots (see
for example in refs.\cite{LBDL98,LBDL98b,KAS93,HEI93,JAC97,MEU92}). In spite
of its simplicity, this model possesses a nonextensive character: it is
inhomogeneous, the total energy does not scale with the particle number of
the system, etc. We will analyze the scaling properties of its asymptotic
accessible volume in order to precise which is its correspondent asymptotic
canonical description in the ThL.

\section{The OLA Model}

The Hamiltonian of this system is given by:

\begin{equation}
H=\stackrel{N}{%
%TCIMACRO{\underset{j=1}{\sum }}%
%BeginExpansion
\mathrel{\mathop{\sum }\limits_{j=1}}%
%EndExpansion
}\frac{1}{2m}{\bf p}_{j}^{2}+\stackrel{N}{%
%TCIMACRO{\underset{j,k=1}{\sum }}%
%BeginExpansion
\mathrel{\mathop{\sum }\limits_{j,k=1}}%
%EndExpansion
}\frac{m\omega ^{2}}{4}\left( {\bf r}_{j}-{\bf r}_{k}\right) ^{2}\text{.}
\label{h}
\end{equation}
This is an example very simple of a {\em self-gravitating small system}{\it %
\ }that allows us a complete analytical study. It will develop its
microcanonical analysis taking also into account the fallowing integrals of
motion:

\begin{equation}
{\bf M}=%
%TCIMACRO{\underset{j=1}{\stackrel{N}{\sum }}}%
%BeginExpansion
\mathrel{\mathop{\stackrel{N}{\sum }}\limits_{j=1}}%
%EndExpansion
{\bf r}_{j}\times {\bf p}_{j}\text{ (angular momentum)},\text{\ \ and\ \ \ }%
{\bf P}_{0}=%
%TCIMACRO{\underset{j=1}{\stackrel{N}{\sum }}}%
%BeginExpansion
\mathrel{\mathop{\stackrel{N}{\sum }}\limits_{j=1}}%
%EndExpansion
{\bf p}_{j}\text{ \ (lineal momentum).}  \label{m}
\end{equation}
In order to eliminate the divergence in the accessible states density of
this model due to the unbound movement of the system mass center, it is also
demand the following additional constrains:

\begin{equation}
{\bf P}_{0}=%
%TCIMACRO{\underset{j=1}{\stackrel{N}{\sum }}}%
%BeginExpansion
\mathrel{\mathop{\stackrel{N}{\sum }}\limits_{j=1}}%
%EndExpansion
{\bf p}_{j}=0\text{\ \ and \ \ }{\bf R}_{0}=%
%TCIMACRO{\underset{j=1}{\stackrel{N}{\sum }}}%
%BeginExpansion
\mathrel{\mathop{\stackrel{N}{\sum }}\limits_{j=1}}%
%EndExpansion
{\bf r}_{j}=0\text{,}  \label{e3}
\end{equation}
which are consistent with the energy and angular momentum conservation. It
is convenient to work with dimensionless variables, so that, it is
introduced the following units:

\begin{equation}
\left[ E\right] =\hbar \omega \text{,}\qquad \left[ M\right] =\hbar \text{,}%
\qquad \left[ R\right] =\left( \frac{\hbar }{m\omega }\right) ^{\frac{1}{2}}%
\text{ and \ }\left[ P\right] =\left( m\hbar \omega \right) ^{\frac{1}{2}}%
\text{.}
\end{equation}

The calculations will be facilitated using the {\em vectorial} {\em %
convention} for the ${\bf R}^{3N}$ vectorial space, which appears in the
appendices. It is easy to see that the equations Eq.(\ref{h}), Eq.(\ref{m})
and Eq.(\ref{e3}) are rewritten as:

\begin{equation}
H=\frac{1}{2}P^{2}+\frac{1}{2}\left( B^{2}R^{2}-\left( B\cdot R\right)
^{2}\right) \text{,}\ \text{\ \ }\ {\bf M}=R\otimes P\ \text{,}\ 
\end{equation}

\begin{equation}
{\bf R}_{0}=B\cdot R=0\ \ \text{and}\ \ {\bf P}_{0}=B\cdot P=0\text{ ,}
\end{equation}
where $R,P\in {\bf R}^{3N}$ are the extended system coordinates and linear
momentum that together conform the {\it N-}body phase space of the system,
and $B\in {\bf R}^{N}$ $\ $with $B=\left( 1,1,\ldots 1\right) $.\ The
solution of this problem can be easily found through of the {\em generating
functional} of the distribution:

\begin{equation}
K\left( \chi ,\upsilon ;I,N\right) =\frac{1}{\hbar ^{7}\omega }\int \frac{%
d^{3N}Rd^{3N}P}{\left( 2\pi \right) ^{3N}}\exp \left[ i\left( \chi \cdot
R+\upsilon \cdot P\right) \right] \delta \left[ I-I_{N}\left( R,P\right) %
\right] \text{,}  \label{e5}
\end{equation}
where $I=\left( E,{\bf M},{\bf R}_{o},{\bf P}_{o}\right) $, $\chi $ and $%
\upsilon $ are {\em 3N-}dimensional vectors belonging to ${\bf R}^{3N}$ . It
is convenient to use the Fourier's representation of the delta function:

\begin{equation}
\delta \left( x_{o}-x\right) =\int_{-\infty }^{+\infty }\frac{dk}{2\pi }\exp %
\left[ z(x_{o}-x)\right] \text{,}
\end{equation}
where $z=\varepsilon +ik$ with $\varepsilon >0$, to rewrite the Eq.(\ref{e5}%
) as follow:

\begin{equation}
K\left( \chi ,\upsilon ;I,N\right) =\frac{1}{\hbar ^{7}\omega }\int \frac{%
d^{10}{\bf k}}{\left( 2\pi \right) ^{10}}\exp \left( {\bf z}\cdot I\right)
\aleph \left( {\bf z};\chi ,\upsilon \right) \text{,}  \label{e6}
\end{equation}
\ where ${\bf k}=\left( k_{1},{\bf k}_{2}{\bf ,}i{\bf \rho ,}i{\bf \eta }%
\right) $, and ${\bf z}=\left( z_{1},{\bf z}_{2}{\bf ,}i{\bf \rho ,}i{\bf %
\eta }\right) $, with $z_{1}=\beta +ik_{1}$, ${\bf z}_{2}={\bf \gamma }%
+ik_{2}$. The function $\aleph \left( {\bf z};\chi ,\upsilon \right) $ is
defined by:

\begin{equation}
\aleph \left( {\bf z};\chi ,\upsilon \right) =\int \frac{d^{3N}Rd^{3N}P}{%
\left( 2\pi \right) ^{3N}}\exp \left\{ -\left[ {\bf z}\cdot I_{N}\left(
R,P\right) -i\chi \cdot R-i\upsilon \cdot P\right] \right\} \text{.}
\label{e7}
\end{equation}
\ 

The above integral is very easy to calculate because it can be reduced to
gaussian integrals. The calculations yield:

\[
K\left( \chi ,\upsilon ;E,{\bf M},N\right) =\frac{1}{\left( 2\pi \right)
^{3}B^{6}\hbar ^{7}\omega }\stackrel{+\infty }{%
%TCIMACRO{\underset{-\infty }{\int }}%
%BeginExpansion
\mathrel{\mathop{\int }\limits_{-\infty }}%
%EndExpansion
}\stackrel{+\infty }{%
%TCIMACRO{\underset{-\infty }{\int }}%
%BeginExpansion
\mathrel{\mathop{\int }\limits_{-\infty }}%
%EndExpansion
}\stackrel{+\infty }{%
%TCIMACRO{\underset{-\infty }{\int }}%
%BeginExpansion
\mathrel{\mathop{\int }\limits_{-\infty }}%
%EndExpansion
}\stackrel{+\infty }{%
%TCIMACRO{\underset{-\infty }{\int }}%
%BeginExpansion
\mathrel{\mathop{\int }\limits_{-\infty }}%
%EndExpansion
}\frac{dk_{1}d^{3}{\bf k}_{2}}{\left( 2\pi \right) ^{4}}\frac{\exp \left(
z_{1}E\right) \exp \left( {\bf z}_{2}\cdot {\bf M}\right) }{\left(
Bz_{1}\right) ^{N-1}\left( B^{2}z_{1}^{2}-z_{2}^{2}\right) ^{N-1}} 
\]

\begin{equation}
\times \exp \left( -\frac{B^{2}\upsilon ^{2}-\left( B\cdot \upsilon \right)
^{2}}{2B^{2}z_{1}}\right) \exp \left( -\frac{B^{2}{\bf z}_{2}^{2}\lambda
^{2}-\left( B{\bf z}_{2}\cdot \lambda \right) ^{2}}{2B^{4}{\bf z}%
_{2}^{2}z_{1}}\right) \exp \left( -\frac{B^{2}\left( {\bf z}_{2}\times
\lambda \right) ^{2}-\left( B{\bf z}_{2}\otimes \lambda \right) ^{2}}{2B^{2}%
{\bf z}_{2}^{2}\left( B^{2}z_{1}^{2}-z_{2}^{2}\right) }z_{1}\right)
\label{e8}
\end{equation}
where:

\begin{equation}
\lambda =\chi +\frac{{\bf z}_{2}\times \upsilon }{z_{1}}
\end{equation}

The above expression must be used in the obtaining of the physical
observables. The accessible states density is expressed as:

\begin{equation}
\Omega \left( E,{\bf M},N\right) =\frac{1}{\left( 2\pi \right)
^{3}B^{6}\hbar ^{7}\omega }\stackrel{+\infty }{%
%TCIMACRO{\underset{-\infty }{\int }}%
%BeginExpansion
\mathrel{\mathop{\int }\limits_{-\infty }}%
%EndExpansion
}\stackrel{+\infty }{%
%TCIMACRO{\underset{-\infty }{\int }}%
%BeginExpansion
\mathrel{\mathop{\int }\limits_{-\infty }}%
%EndExpansion
}\stackrel{+\infty }{%
%TCIMACRO{\underset{-\infty }{\int }}%
%BeginExpansion
\mathrel{\mathop{\int }\limits_{-\infty }}%
%EndExpansion
}\stackrel{+\infty }{%
%TCIMACRO{\underset{-\infty }{\int }}%
%BeginExpansion
\mathrel{\mathop{\int }\limits_{-\infty }}%
%EndExpansion
}\frac{dk_{1}d^{3}{\bf k}_{2}}{\left( 2\pi \right) ^{4}}\frac{\exp \left(
z_{1}E\right) \exp \left( {\bf z}_{2}\cdot {\bf M}\right) }{\left(
Bz_{1}\right) ^{N-1}\left( B^{2}z_{1}^{2}-z_{2}^{2}\right) ^{N-1}}
\label{sd}
\end{equation}
whose integration yields:

\begin{equation}
\Omega \left( E,{\bf M},N\right) =\frac{1}{\left( 2\pi \hbar B\right)
^{7}\omega }M^{3N-7}H_{N}\left( \frac{E}{BM}\right) \text{. }
\end{equation}
The function $H_{N}\left( x\right) $ is given by:

\begin{equation}
H_{N}\left( x\right) =\left( 1-x\right) ^{2N-4}P_{N-2}\left( 1-x\right)
\sigma \left( x-1\right) ,\text{ where\ }\sigma \left( x\right) =\left\{ 
\begin{array}{c}
1\text{ if }x>0 \\ 
0\text{ if }x<0
\end{array}
\right. \text{ is the Heaviside function,}
\end{equation}
\begin{equation}
\text{and }P_{N-2}(z)=\stackrel{N-2}{%
%TCIMACRO{\underset{n=0}{\sum }}%
%BeginExpansion
\mathrel{\mathop{\sum }\limits_{n=0}}%
%EndExpansion
}a_{n}z^{n}\text{, \ with \ \ }a_{n}=\frac{\left( N-1+n\right) !}{\left(
N-2-n\right) !\left( 2N-4+n\right) !\left( N-1\right) !}\frac{\left(
-1\right) ^{n}}{2^{n}n!}\text{.\ }  \label{pd}
\end{equation}

The accessible volume is obtained multiplying the state density by an
appropriate element volume constant $\delta I$. A reasonable choice for the
tridimensional OLA model is:

\begin{equation}
\delta I=\hbar ^{7}\omega \text{,}
\end{equation}
and therefore:

\[
W\left( E,{\bf M},N\right) =\frac{1}{\left( 2\pi B\right) ^{7}}%
M^{3N-7}H_{N}\left( \frac{E}{BM}\right) \text{.} 
\]
The integrals of motion space, $\Im _{N}$, in the ${\cal R}_{\left( E,{\bf M,%
}N\right) }$ representation is the ${\bf R}^{4}$ cone for a given $N$:

\begin{equation}
E^{2}-N{\bf M}^{2}\geq 0\text{.}
\end{equation}

In order to access to the scaling laws, it must be obtained the asymptotic
dependency of the polynomials coefficients in the Eq.(\ref{pd}). It is easy
to show the following asymptotic behavior:

\begin{equation}
%TCIMACRO{\underset{N\rightarrow \infty }{\lim }}%
%BeginExpansion
\mathrel{\mathop{\lim }\limits_{N\rightarrow \infty }}%
%EndExpansion
a_{n}\simeq \frac{1}{2^{2N}N^{3N}}\frac{\left( -1\right) ^{n}N^{n}}{4^{n}n!}%
\text{.}
\end{equation}
Therefore, the asymptotic accessible volume is given by the expression:

\begin{equation}
W_{asym}\left( E,{\bf M},N\right) \propto \left( \frac{E}{2\left( N\right) ^{%
\frac{3}{2}}}\right) ^{3N}Q_{N}\left( \frac{\sqrt{N}M}{E}\right) \text{,}
\label{aw}
\end{equation}
where the function $Q_{N}\left( x\right) $ is expressed as:

\begin{equation}
Q_{N}\left( x\right) =\left( 1-x\right) ^{2N}\exp \left( N\frac{1-x}{4}%
\right) \text{, \ with \ }x\in \left[ 0,1\right] \text{.}
\end{equation}

It is very easy to see that the tridimensional OLA model possesses an {\em %
exponential self-similarity scaling laws }\cite{vel1,vel2}:

\begin{equation}
\left. 
\begin{array}{c}
N_{o}\rightarrow N\left( \alpha \right) =\alpha N_{o} \\ 
E_{o}\rightarrow E\left( \alpha \right) =\alpha ^{\frac{3}{2}}E_{o} \\ 
{\bf M}_{o}\rightarrow {\bf M}\left( \alpha \right) =\alpha {\bf M}_{o}
\end{array}
\right\} \Rightarrow W_{asym}\left( \alpha \right) ={\cal F}\left[
W_{o},\alpha \right] \text{,}
\end{equation}
where the functional ${\cal F}\left[ W_{o},\alpha \right] $ is given by:

\begin{equation}
{\cal F}\left[ W_{o},\alpha \right] \equiv \exp \left[ \alpha \ln W_{o}%
\right] \text{.}
\end{equation}
where $\alpha $ is the scaling parameter and $W_{o}$ is given by:

\begin{equation}
W_{o}=W_{asym}\left( E_{o},{\bf M}_{o},N_{o}\right) \text{.}
\end{equation}
Thus, the tridimensional OLA model is not extensive, but {\em %
pseudoextensive }\cite{vel2}. In the present case, the generalized
Boltzmann's Principle \cite{vel1} adopts its ordinary form:

\begin{equation}
\begin{tabular}{|l|}
\hline
$S_{B}=\ln W,$ \\ \hline
\end{tabular}
\end{equation}
his celebrated gravestone epitaph in Vienna. To this entropy form
corresponds the usual Shannon-Boltzmann-Gibbs' extensive entropy:

\begin{equation}
S_{SBG}=-%
%TCIMACRO{\underset{k}{\sum }}%
%BeginExpansion
\mathrel{\mathop{\sum }\limits_{k}}%
%EndExpansion
p_{k}\ln p_{k}\text{,}
\end{equation}
and therefore, in the thermodynamic limit, the microcanonical description
could be substituted equivalently by the usual Boltzmann-Gibbs' Distribution 
{\em with a appropriate selection of the representation of the integrals of
motion space}. This can be performed changing the representation from ${\cal %
R}_{\left( E,{\bf M},N\right) }$ to ${\cal R}_{\left( {\cal E},{\bf M}%
,N\right) }$, where:

\begin{equation}
{\cal E}=\frac{1}{\sqrt{N}}E\text{.}
\end{equation}

In the thermodynamic limit the microcanonical description of this system
should be substituted equivalently by the canonical one. In this case the
Boltzmann-Gibbs' Distribution is:

\begin{equation}
\omega _{c}\left( X;\beta ,{\bf \gamma },N\right) =\frac{1}{Z\left( \beta ,%
{\bf \gamma },N\right) }\exp \left[ -\beta {\cal E}_{N}\left( X\right) -{\bf %
\gamma }\cdot {\bf M}_{N}\left( X\right) \right] \text{.}  \label{bgd1}
\end{equation}

Through the Laplace's Transformation:

\begin{equation}
Z\left( \beta ,{\bf \gamma },N\right) =\int \exp \left( -\beta {\cal E}-{\bf %
\gamma }\cdot {\bf M}\right) W\left( {\cal E},{\bf M},N\right) \frac{d{\cal E%
}d^{3}{\bf M}}{\delta I}\text{,}  \label{tl}
\end{equation}
it is easy to show that in the thermodynamic limit both descriptions are
equivalent, that is, the {\em Legendre's Transformation} between the
thermodynamic potentials of the ensembles is valid:

\begin{equation}
P\left( \beta ,{\bf \gamma },N\right) \simeq \beta {\cal E}+{\bf \gamma }%
\cdot {\bf M}-S_{B}\left( {\cal E},{\bf M},N\right) \text{,}
\end{equation}
where $P\left( \beta ,{\bf \gamma },N\right) $ is the {\em Planck's Potential%
}:

\begin{equation}
P\left( \beta ,{\bf \gamma },N\right) =-\ln Z\left( \beta ,{\bf \gamma }%
,N\right) \text{,}
\end{equation}
when exists {\em a unique sharp maximum} in the integral argument of the Eq.(%
\ref{tl}). In this maximum, $\left( {\cal E}_{m},{\bf M}_{m}\right) $, it
will be satisfied the following conditions:

\begin{equation}
\beta =\frac{\partial }{\partial {\cal E}}S_{B}\left( {\cal E}_{m},{\bf M}%
_{m},N\right) \text{, }{\bf \gamma =}\frac{\partial }{\partial {\bf M}}%
S_{B}\left( {\cal E}_{m},{\bf M}_{m},N\right) \text{,}
\end{equation}
and all the eingenvalues of the {\em curvature tensor }\cite{gro1,gro2}:

\begin{equation}
K_{\mu \nu }=\partial _{\mu }\partial _{\nu }S_{B}\left( {\cal E}_{m},{\bf M}%
_{m},N\right) \text{,}  \label{ct}
\end{equation}
are negatives [in the Eq.(\ref{ct}) $\partial _{\mu }\equiv \frac{\partial }{%
\partial I^{\mu }}$, where $I^{\mu }=\left( {\cal E},{\bf M}\right) $].
Similarly, the generalized Duhem-Gibbs' relation is also valid in the
thermodynamic limit:

\begin{equation}
S_{B}\left( {\cal E},{\bf M},N\right) =\beta {\cal E}+{\bf \gamma }\cdot 
{\bf M+}\mu N\text{,}
\end{equation}
where $\mu $ is the chemical potential:

\begin{equation}
\mu \left( \beta ,{\bf \gamma }\right) =\frac{\partial }{\partial N}%
S_{B}\left( {\cal E}_{m},{\bf M}_{m},N\right) .
\end{equation}
From the Duhem-Gibbs' relation is easily deduced the following relationship:

\begin{equation}
P\left( \beta ,{\bf \gamma },N\right) =-\mu \left( \beta ,{\bf \gamma }%
\right) N\text{.}
\end{equation}

The Boltzmann's entropy of the tridimensional OLA model in the asymptotic
region is given by:

\begin{equation}
S_{B}\simeq 3N\ln \left( \frac{{\cal E}}{2N}\right) +2N\ln \left( 1-\frac{M}{%
{\cal E}}\right) +\frac{1}{4}N\left( 1-\frac{M}{{\cal E}}\right) +O\left( 
\frac{1}{N}\right) \text{,}
\end{equation}
and therefore, the canonical parameters $\beta \ $and $\gamma $, and the
chemical potential $\mu $, are given by:

\begin{equation}
\beta =\frac{1}{4}N\frac{12{\cal E}^{2}-3{\cal E}M-M^{2}}{{\cal E}^{2}\left( 
{\cal E}-M\right) },\text{ }{\bf \gamma }=\allowbreak -\frac{1}{4}N\frac{9%
{\cal E}-M}{{\cal E}\left( {\cal E}-M\right) }\frac{{\bf M}}{M},\text{ }\mu =%
\frac{S_{B}\left( {\cal E},{\bf M},N\right) }{N}-3.
\end{equation}
To validate the previous results it must be demanded the concavity of the
Boltzmann's entropy, that is to say, the non-negativity of the $4\times 4$
curvature tensor:

\begin{equation}
K_{\mu \nu }=\left( 
\begin{tabular}{ll}
$-\frac{1}{2}N\frac{6{\cal E}^{3}-3{\cal E}^{2}M+M^{3}}{{\cal E}^{3}\left( 
{\cal E}-M\right) ^{2}}$ & $\ \ \ \ \ \ \ \ \ \ \ \ \ \ \ \frac{1}{4}N\frac{9%
{\cal E}^{2}-2{\cal E}M+M^{2}}{{\cal E}^{2}\left( {\cal E}-M\right) ^{2}}%
\frac{{\bf M}_{3\times 1}}{M}$ \\ 
$\frac{1}{4}N\frac{9{\cal E}^{2}-2{\cal E}M+M^{2}}{{\cal E}^{2}\left( {\cal E%
}-M\right) ^{2}}\frac{{\bf M}_{3\times 1}}{M}$ & $\ \ \ -N\frac{2}{\left( 
{\cal E}-M\right) ^{2}}\frac{{\bf M}_{3\times 1}}{M}\frac{{\bf M}_{1\times 3}%
}{M}-\frac{1}{4}N\frac{9{\cal E}-M}{{\cal E}\left( {\cal E}-M\right) M}%
\left( {\bf I}_{tr}\right) _{3\times 3}$%
\end{tabular}
\right)
\end{equation}
where ${\bf M}_{1\times 3}=\left( {\bf M}_{3\times 1}\right) ^{T}=\left(
M_{x},M_{y},M_{z}\right) $, and ${\bf I}_{tr}$ is the $3\times 3$ transverse
unitary matrix:

\begin{equation}
\left( {\bf I}_{tr}\right) _{\mu \nu }=\delta _{\mu \nu }-\frac{M_{\mu }}{M}%
\frac{M_{\nu }}{M}\text{.}
\end{equation}
The determinant of the curvature tensor is given by:

\begin{equation}
\det K=\frac{1}{256}N^{4}\frac{15{\cal E}^{2}+18{\cal E}M-M^{2}}{\left( 
{\cal E}-M\right) ^{4}{\cal E}^{6}M^{2}}\left( 9{\cal E}-M\right) ^{2}>0.
\label{det}
\end{equation}

From the above result is derived that all the accessible space is
appropriately described by the canonical ensemble in the ${\cal R}_{\left( 
{\cal E},{\bf M},N\right) }$ representation: in the Laplace's
Transformation, Eq.(\ref{tl}), there is only one sharp peak. Thus, in the
thermodynamic limit \ both descriptions are identical: all the system
accessible space can be deal with the canonical description. This result
facilitates so much the analysis of this system. In the canonical
description, the Planck's potential is obtained from the Eq.(\ref{e8})
replacing $Bz_{1}\rightarrow \beta $ and ${\bf z}_{2}\rightarrow {\bf \gamma 
}$:

\begin{equation}
P\left( \beta ,{\bf \gamma },N\right) =N\ln \left[ \beta \left( \beta
^{2}-\gamma ^{2}\right) \right] \text{.}
\end{equation}

The generating functional in the canonical description:

\begin{equation}
G_{c}\left( \chi ,\upsilon ;\beta ,{\bf \gamma },N\right) =\int \frac{%
d^{3N}Qd^{3N}P}{\left( 2\pi \right) ^{3N}}\omega _{c}\left( Q,P;\beta ,{\bf %
\gamma },N\right) \exp \left( i\chi \cdot Q+i\upsilon \cdot P\right) \text{,}
\end{equation}
is also obtained from the Eq.(\ref{e8}):

\begin{equation}
G_{c}\left( \chi ,\upsilon ;\beta ,{\bf \gamma },N\right) =\exp \left( -%
\frac{B^{2}\upsilon ^{2}-\left( B\cdot \upsilon \right) ^{2}}{2B\beta }%
\right) \exp \left( -\frac{B^{2}{\bf \gamma }^{2}\lambda ^{2}-\left( B{\bf %
\gamma }\cdot \lambda \right) ^{2}}{2B^{3}{\bf \gamma }^{2}\beta }\right)
\exp \left( -\frac{B^{2}\left( {\bf \gamma }\times \lambda \right)
^{2}-\left( B{\bf \gamma }\otimes \lambda \right) ^{2}}{2B^{3}{\bf \gamma }%
^{2}\left( \beta ^{2}-{\bf \gamma }^{2}\right) }\beta \right) \text{.}
\end{equation}

It is very interesting to known the particle distribution in the space. This
observable is easily obtained from the generating functional as follow:
setting zero all the components of the vectors $\upsilon $ and leaving only
one vectorial component in the vector $\chi $, $\chi =({\bf q},0,....,0)$,
and performing the following integration: 
\begin{equation}
\rho \left( {\bf r}\right) =N\stackrel{+\infty }{%
%TCIMACRO{\underset{-\infty }{\int }}%
%BeginExpansion
\mathrel{\mathop{\int }\limits_{-\infty }}%
%EndExpansion
}\stackrel{+\infty }{%
%TCIMACRO{\underset{-\infty }{\int }}%
%BeginExpansion
\mathrel{\mathop{\int }\limits_{-\infty }}%
%EndExpansion
}\stackrel{+\infty }{%
%TCIMACRO{\underset{-\infty }{\int }}%
%BeginExpansion
\mathrel{\mathop{\int }\limits_{-\infty }}%
%EndExpansion
}\frac{dq_{l}d^{2}{\bf q}_{t}}{\left( 2\pi \right) ^{3}}\exp \left(
-ir_{l}q_{l}\right) \exp \left( -i{\bf r}_{t}\cdot {\bf q}_{t}\right) \exp
\left( -\frac{B^{2}-1}{2B^{3}\beta }q_{l}^{2}\right) \exp \left( -\frac{%
B^{2}-1}{2B^{3}\left( \beta ^{2}-{\bf \gamma }^{2}\right) }\beta {\bf q}%
_{t}^{2}\right) \text{,}
\end{equation}
where $q_{l}$ and ${\bf q}_{t}$ are the longitudinal and the transverse
components of the vector ${\bf q}$ with respect to the vector ${\bf \gamma }$%
. The integration yields:

\begin{equation}
\rho \left( {\bf r}\right) =\frac{N}{\sqrt{8\pi ^{3}\sigma _{l}\sigma
_{t}^{2}}}\exp \left( -\frac{r_{l}^{2}}{2\sigma _{l}}-\frac{{\bf r}_{t}^{2}}{%
2\sigma _{t}}\right) ,
\end{equation}
showing that the particle distribution possesses a gaussian shape where the
parameters $\sigma _{l}$ and $\sigma _{t}$ are given by:

\begin{equation}
\sigma _{l}=\frac{B^{2}-1}{B^{3}\beta }\text{ , }\sigma _{t}=\frac{B^{2}-1}{%
B^{3}\left( \beta ^{2}-{\bf \gamma }^{2}\right) }\beta \text{.}  \label{sp}
\end{equation}

Let the parameter $s$ be introduced as follow:

\begin{equation}
\frac{r_{l}^{2}}{\left( \frac{\sigma _{l}}{\sigma }\right) s^{2}}+\frac{{\bf %
r}_{t}^{2}}{\left( \frac{\sigma _{t}}{\sigma }\right) s^{2}}=1,
\end{equation}
parametrizing a family of ellipsoids of revolution which characterizing the
surfaces of equal density:

\begin{equation}
\rho \left( s\right) =\frac{N}{\sqrt{8\pi ^{3}\sigma ^{3}}}\exp \left( -%
\frac{s^{2}}{2\sigma }\right) ,
\end{equation}
where $\sigma $ is defined by the relation:

\begin{equation}
\sigma =\sqrt[3]{\sigma _{l}\sigma _{t}^{2}}.
\end{equation}
The eccentricity of ellipsoids family is constant and given by:

\begin{equation}
e=\sqrt{1-\frac{\sigma _{l}}{\sigma _{t}}}=\frac{\gamma }{\beta }=\frac{M}{%
{\cal E}}\text{.}
\end{equation}

As it can be seen, the rotation deforms the spherical shape of the
distribution. The characteristic size of the distribution is the given by:

\begin{equation}
l_{c}=\sqrt{\sigma }.
\end{equation}
From the Eq.(\ref{sp}) it is deduced that the characteristic size decreases
with the increasing of the particle number $N$:

\begin{equation}
l_{c}\sim 1/\sqrt[4]{N}\text{.}
\end{equation}
This kind of scaling law is not typical in the extensive systems, which are
usually scaled proportional to the systems size. In the FIG.1. it is shown
different profiles for the particle distributions.

\section{Conclusions}

In the present paper we applied the results of our previous works on study
of a very simple model of self-gravitating small system, the tridimensional
OLA model. The microcanonical analysis showed that the OLA model is a
pseudoextensive system \cite{vel2}, since it possesses a exponential
self-similarity scaling laws in the thermodynamic limit. Systems with this
kind of scaling laws could be dealt with the usual Boltzmann-Gibbs'
statistics if an appropriate representation of the system integrals of
motion space is selected. Although the model presented before is very
simple, the analysis shown that it is necessary the consideration of
self-similarity scaling postulates for the analysis of this system, since
the same is not extensive.

\section{Appendix: The vectorial convention.}

Let ${\bf R}^{3N}$\ be the $3N$-dimensional vectorial space, which is the
external product of the spaces ${\bf R}^{3}$, ${\bf R}^{N}$. The bases of \ $%
{\bf R}^{3N}$\ are represented by means of the product of the bases of the
respective spaces, ${\bf R}^{3}$\ and ${\bf R}^{N}$: 
\begin{equation}
\overrightarrow{E}_{k,s}=\overrightarrow{e}_{s}\overrightarrow{E}_{k}\text{ }%
.
\end{equation}
\ It will be only considered those transformations preserving this property,
that is, those transformations which acts in the spaces ${\bf R}^{3}$\ and $%
{\bf R}^{N}$\ separately. The vectors of this space can be represented as: 
\begin{equation}
V=\stackrel{N}{%
%TCIMACRO{\underset{k=1}{\sum }}%
%BeginExpansion
\mathrel{\mathop{\sum }\limits_{k=1}}%
%EndExpansion
}\stackrel{3}{%
%TCIMACRO{\underset{s=1}{\sum }}%
%BeginExpansion
\mathrel{\mathop{\sum }\limits_{s=1}}%
%EndExpansion
}X_{k,s}\overrightarrow{e}_{s}\overrightarrow{E}_{k}=\stackrel{N}{%
%TCIMACRO{\underset{k=1}{\sum }}%
%BeginExpansion
\mathrel{\mathop{\sum }\limits_{k=1}}%
%EndExpansion
}\overrightarrow{x}_{k}\overrightarrow{E}_{k}=\stackrel{3}{%
%TCIMACRO{\underset{s=1}{\sum }}%
%BeginExpansion
\mathrel{\mathop{\sum }\limits_{s=1}}%
%EndExpansion
}\overrightarrow{X}_{s}\overrightarrow{e}_{s}\text{ \ \ where \ }%
\overrightarrow{x}_{k}\in {\bf R}^{3};\overrightarrow{X}_{s}\in {\bf R}^{N}%
\text{ \ }.
\end{equation}

From here and after, it will be adopted the tensorial summation convention.
Using the above representation, it can be defined the following external
operations:

Let be the vectors: $\overrightarrow{a}\in {\bf R}^{3};\overrightarrow{A}\in 
{\bf R}^{N};\overrightarrow{V}_{1},\overrightarrow{V}_{2},\overrightarrow{V}%
_{3}\in {\bf R}^{3N}$.

\begin{center}
{\bf Partial and Total Inner Operations:} 
\begin{equation}
R^{3}\times R^{3N}\rightarrow R^{N}:\overrightarrow{a}\cdot \overrightarrow{V%
}=\left( \overrightarrow{a}\cdot \overrightarrow{v}_{k}\right) 
\overrightarrow{E}_{k},
\end{equation}
\begin{equation}
{\bf R}^{N}\times {\bf R}^{3N}\rightarrow {\bf R}^{3}:\overrightarrow{A}%
\cdot \overrightarrow{V}=\left( \overrightarrow{A}\cdot \overrightarrow{V}%
_{s}\right) \overrightarrow{e}_{s},
\end{equation}
\begin{equation}
{\bf R}^{3}\times {\bf R}^{3N}\rightarrow {\bf R}^{N}:\overrightarrow{V}%
\cdot \overrightarrow{W}=\left( \overrightarrow{v_{k}}\cdot \overrightarrow{w%
}_{k}\right) =\left( \overrightarrow{V_{s}}\cdot \overrightarrow{W}%
_{s}\right) \text{.}
\end{equation}

{\bf Partial Vectorial Product:} 
\begin{equation}
{\bf R}^{3}\times {\bf R}^{3N}\rightarrow {\bf R}^{3N}:\overrightarrow{a}%
\times \overrightarrow{V}=\left( \overrightarrow{a}\times \overrightarrow{v}%
_{k}\right) \overrightarrow{E}_{k}\text{ .}
\end{equation}

{\bf Vectorial-Scalar Product:} 
\begin{equation}
{\bf R}^{3N}\otimes {\bf R}^{3N}\rightarrow {\bf R}^{3}:\overrightarrow{V}%
\otimes \overrightarrow{W}=\left( \overrightarrow{v_{k}}\times 
\overrightarrow{w}_{k}\right) \text{.}
\end{equation}
\end{center}

From the previous definitions is derived the following identity:

\begin{equation}
\overrightarrow{a}\cdot \overrightarrow{V}\otimes \overrightarrow{W}=\left( 
\overrightarrow{a}\times \overrightarrow{V}\right) \cdot \overrightarrow{W}\
.
\end{equation}

\ \ \ \ \ \ \ \ \ \ \ \

\begin{center}
\bigskip
\end{center}


\begin{references}
\bibitem{vel1}  L. Velazquez and F. Guzman, Generalizing the Extensive
Postulates, preprint (2001) [cond-mat/0107214].

\bibitem{vel2}  L. Velazquez and F. Guzman, Microcanonical theory and the
pseudoextensive systems; preprint (2001) [cond-mat/0107439].

\bibitem{gro1}  D. H. E Gross, {\it Microcanonical thermodynamics: Phase
transitions in Small systems}, {\it 66 Lectures Notes in Physics}, (World
scientific, Singapore, 2001) and refs. therein.

\bibitem{gro2}  D. H. E. Gross, Chaos, Solitons, and Fractals13 (2001);
preprint (2000) [cond-mat/0004268].

\bibitem{LBDL98}  F. Luczak, F. Brosens, J. T. Devreese, and L. F. Lemmens,
Phys. Rev. E {\bf 57}, 2411 (1998).

\bibitem{LBDL98b}  F. Luczak, F. Brosens, J. T. Devreese, and L. F. Lemmens,
Int. J. Mod. Phys. C., {\bf 10,}1(1999).

\bibitem{KAS93}  M. A. Kastner, Physics Today, January 1993.

\bibitem{HEI93}  D. Heitmann, and J. P. Kotthaus, Physics Today, June 1993.

\bibitem{JAC97}  L. Jacak, P. Hawrylak, and A. Wojs, {\sl Quantum Dots},
Springer 1997.

\bibitem{MEU92}  B. Meurer, D. Heitmann, and K. Ploog, Phys. Rev. Lett. {\bf %
\ 68}, 1371 (1992); Phys. Rev. B {\bf 48}, 11488 (1993).
\end{references}
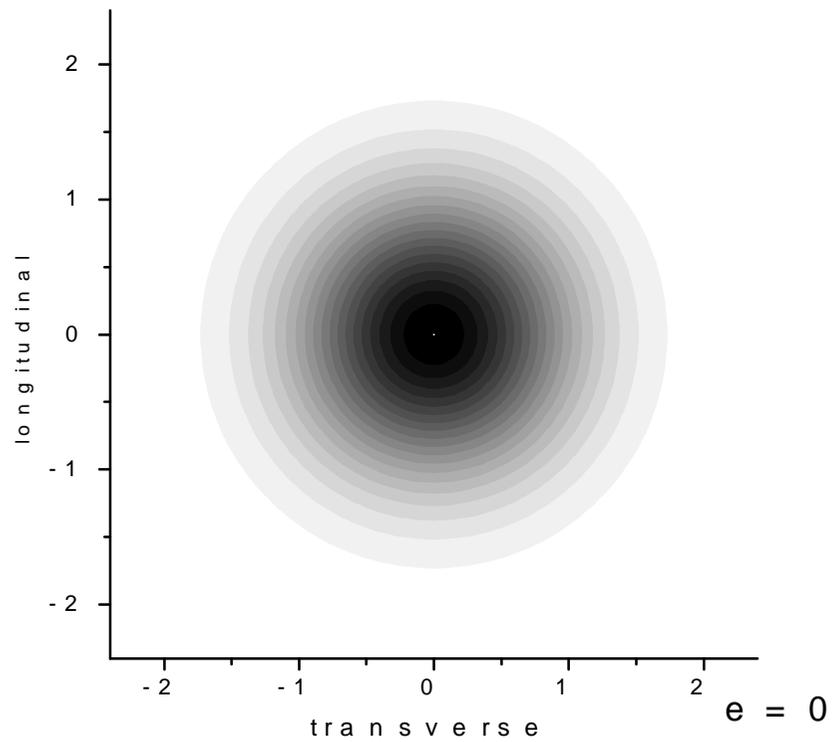
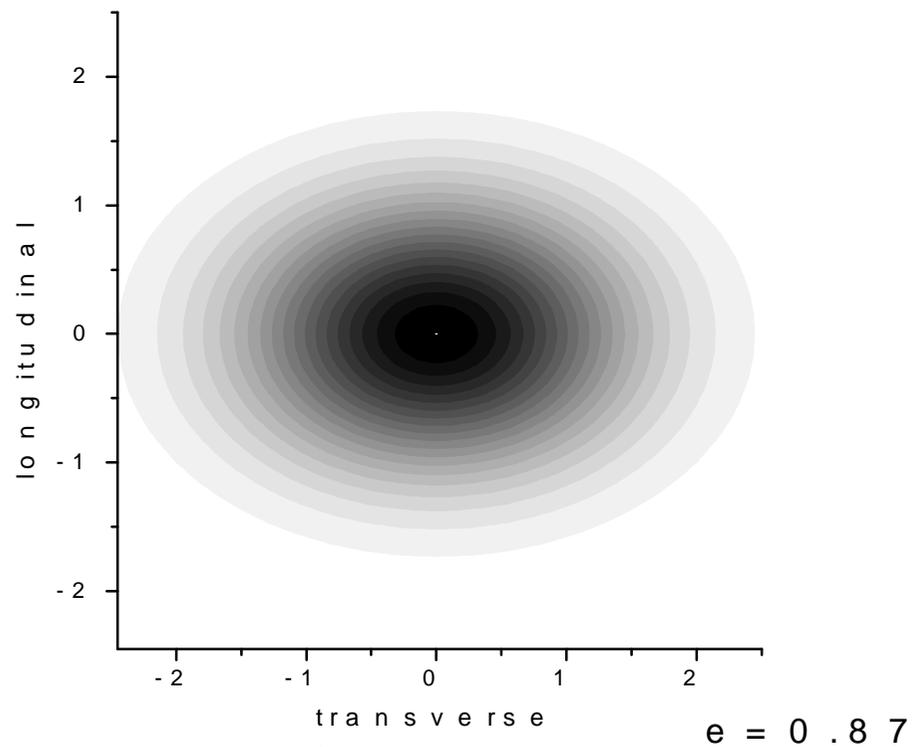
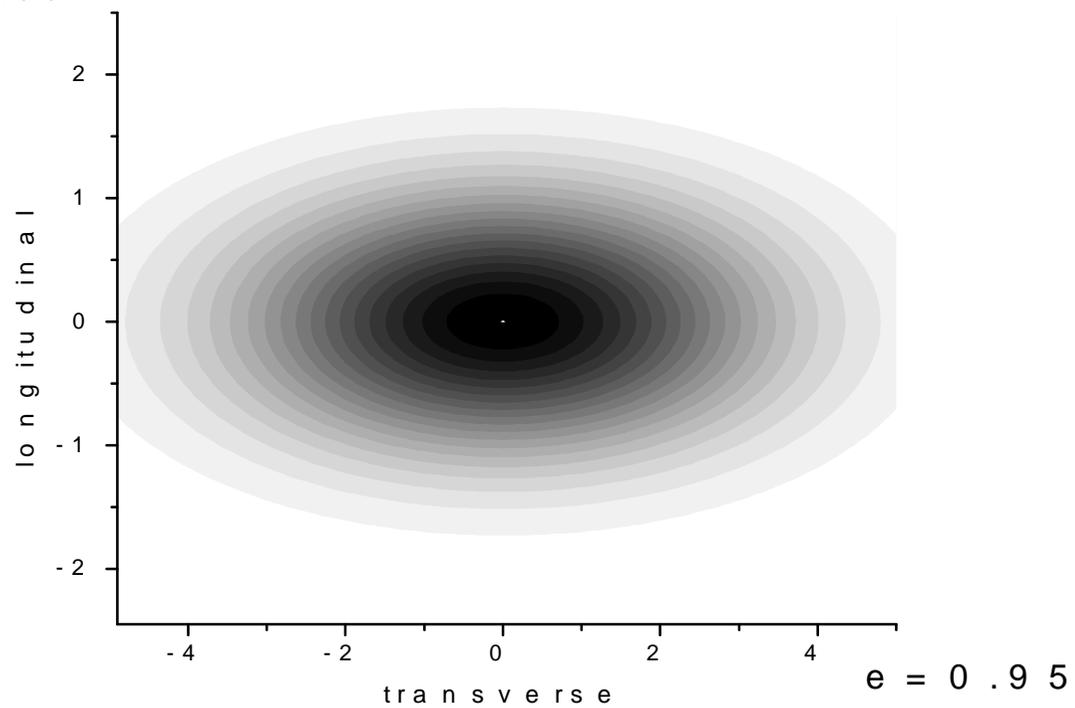
\end{document}